\documentclass[letterpaper,twocolumn,english,amssymb,aps,prb]{revtex4}
\usepackage[T1]{fontenc}
\usepackage[latin9]{inputenc}
\usepackage{units}
\usepackage{amstext}
\usepackage{graphicx}
\usepackage{amssymb}

\makeatletter

\providecommand*{\perispomeni}{\char126}
\AtBeginDocument{\DeclareRobustCommand{\greektext}{%
  \fontencoding{LGR}\selectfont\def\encodingdefault{LGR}%
  \renewcommand{\~}{\perispomeni}%
}}
\DeclareRobustCommand{\textgreek}[1]{\leavevmode{\greektext #1}}
\DeclareFontEncoding{LGR}{}{}

\newcommand{\lyxmathsym}[1]{\ifmmode\begingroup\def\b@ld{bold}
  \text{\ifx\math@version\b@ld\bfseries\fi#1}\endgroup\else#1\fi}

\@ifundefined{textcolor}{}
{%
 \definecolor{BLACK}{gray}{0}
 \definecolor{WHITE}{gray}{1}
 \definecolor{RED}{rgb}{1,0,0}
 \definecolor{GREEN}{rgb}{0,1,0}
 \definecolor{BLUE}{rgb}{0,0,1}
 \definecolor{CYAN}{cmyk}{1,0,0,0}
 \definecolor{MAGENTA}{cmyk}{0,1,0,0}
 \definecolor{YELLOW}{cmyk}{0,0,1,0}
 }

\makeatother

\usepackage{babel}

\begin{document}

\title{Evidence for Helical Edge Modes in Inverted InAs/GaSb Quantum Wells}

\author{Ivan Knez}

\affiliation{Department of Physics and Astronomy, Rice University, Houston, TX
77251-1892}

\author{Rui-Rui Du}

\affiliation{Department of Physics and Astronomy, Rice University, Houston, TX
77251-1892}

\author{Gerard Sullivan}

\affiliation{Teledyne Scientific and Imaging, Thousand Oaks, CA 91630}
\begin{abstract}
We present an experimental study of low temperature electronic transport
in the hybridization gap of inverted InAs/GaSb composite quantum wells.
Electrostatic gate is used to push the Fermi level into the gap regime,
where the conductance as a function of sample length and width is
measured. Our analysis shows strong evidence for the existence of
helical edge modes proposed by Liu et al {[}Phys. Rev. Lett., 100,
236601 (2008){]}. Edge modes persist inspite of sizable bulk conduction
and show only a weak magnetic field dependence - a direct consequence
of gap opening away from zone center.
\end{abstract}
\maketitle


Topological insulators (TI) are a novel phase of matter,\citep{1,2}
originally predicted to manifest in 2D structures\citep{3} as a superposition
of two quantum Hall systems,\citep{4} where the role of the spin-dependent
magnetic field is played by the spin-orbital interactions. In extension
of the paradigm to 3D, TI surfaces emerge as \textquotedbl{}half-graphene\textquotedbl{}
with an odd number of Dirac cones.\citep{5} In 2D, the TI phase is
also known as quantum spin Hall insulator (QSHI) and is characterized
by an energy gap in the bulk and topologically protected helical edge
states. Quantized conductance, taken as the evidence for the QSHI
phase, has been experimentally observed in the inverted HgTe/CdTe
quantum wells (QWs).\citep{6,7} Liu et al\citep{8} have proposed
that QSHI should arise in another semiconductor system, the hybridized
InAs/GaSb QWs, where a rich phase diagram including band insulator
and QSHI can be continuously tuned via gate voltages. Here we present
a systematic transport study of high quality InAs/GaSb devices tuned
into the QSHI state, where we observe slowly-propagating helical edge
modes that are largely immune to a conductive bulk. Exploring this
system should have a far-reaching impact, since InAs makes a good
interface with superconductors,\citep{9} a prerequisite for fabricating
TI/superconductor hybrid structures;\citep{10} the latter are predicted
to host exotic Majorana fermion modes and are viable for fault-tolerant
quantum computing. 

A common characteristic to all TIs is band inversion, which in InAs/GaSb
is achieved by tuning energy levels in two neighboring electron and
hole QWs. Hybridization of electron-hole bands leads to a gap opening,
which has been experimentally well established, albeit always with
a non-zero residual conductivity.\citep{11,12} In an early theoretical
study,\citep{13} the origin of the residual conductivity has been
ascribed to the level-broadening due to scattering. Interestingly,
in the \textquotedblleft{}clean limit\textquotedblright{}, the gap
conductivity is finite, yet independent of scattering parameters,
such as sample mobility. Motivated by the QSHI proposal,\citep{8}
Knez et al\citep{14} revisited the issue of gap conduction in micro-size
samples. They found a bulk conductivity on the order of $\sim10e^{2}/h$,
consistent with\citep{13} and a few times larger than the expected
contribution from the edge. Nevertheless, bulk conductivity diminishes
as the band inversion is reduced,\citep{14} promoting the QSHI. In
this Letter we study the length and width dependence of conductance
in the hybridization regime and find direct evidence for the existence
of helical edge modes proposed by Liu et al.\citep{8} Surprisingly,
edge modes persist alongside the conductive bulk and show only weak
magnetic field dependence. This apparent decoupling between the edge
and bulk is a direct consequence of gap opening away from the zone
center, which leads to a large disparity in Fermi wave-vectors between
bulk and edge states, and results in a qualitatively different QSHI
phase than in HgTe/CdTe where the gap opens at the zone center. 

InAs/GaSb has a broken gap band alignment allowing for the coexistence
of closely separated electron (in InAs) and hole (in GaSb) two-dimensional
gases and confined by neighboring AlSb barriers as shown in Fig. 1a.\citep{15}
For wider wells the band structure is inverted with the ground conduction
subband (E1) lower than the ground heavy-hole subband (H1). Relative
position of the E1 and H1 bands can be tuned by an external electric
field\citep{15} applied via front and back gates. In inverted regime
the E1 and H1 bands anti-cross for some finite wave-vector $k_{cross}$,
where electron and hole densities are approximately matched, $n=p=k_{cross}^{2}/2\pi$.
Due to the tunneling between the wells, electron and hole states are
mixed and a hybridization gap $\lyxmathsym{\textgreek{D}}$ opens
in the otherwise semi-metallic band dispersion as shown in Fig. 1b.\citep{15}
Matching of the inverted bands to the corresponding vacuum states
leads to an inevitable gap closing at the sample perimeter and results
in linearly dispersing edge modes.\citep{8} Time reversal symmetry
of the governing Hamiltonian requires the edge modes to be helical,
i.e. counter-propagating spin up and spin down channels with conserved
helicity. As a result, particles on time-reversed paths around a non-magnetic
impurity in the helical edge destructively interfere, resulting in
zero backscattering probability.\citep{2} For Fermi energy $E_{F}$
in the gap, expected edge conductance in a six-terminal configuration
for mesoscopic samples will be $2e^{2}/h$.\citep{7} Here we use
a four-terminal configuration where expected edge conductance is doubled
to $4e^{2}/h$. 

Longer samples can be modeled by inserting phase breaking probes\citep{16}
and applying the Landauer-Buttiker formula yielding four-terminal
conductance as: 

\begin{equation}
G_{14,23}=\frac{2e^{2}}{h}\left(\frac{l_{\phi}}{L}+\left(\frac{l_{\phi}}{L}\right)^{2}\right).\label{eq:1}\end{equation}
where $l_{\phi}$ is the phase coherence length and $L$ is the device
length. Thus, for macroscopic QSH samples $\left(L\gg l_{\phi}\right)$
the edge contribution to the conductance will be negligible. Note
that due to the level broadening $\Gamma$ the hybridization gap exhibits
a sizable bulk conductivity, which for small level broadening $\Gamma\ll\Delta$,
scales as $g_{bulk}\sim\frac{e^{2}}{h}\frac{E_{g0}}{\Delta}$,\citep{13,14}
where $E_{g0}$ is the relative separation between H1 and E1 bands.
While helical edge transport manifests itself only in the mesoscopic
regime, macroscopic samples can be used as an important diagnostic
of bulk gap conduction, allowing us to separate edge from bulk contributions
which coexist in mesoscopic samples.

\begin{figure}
\includegraphics[scale=0.9]{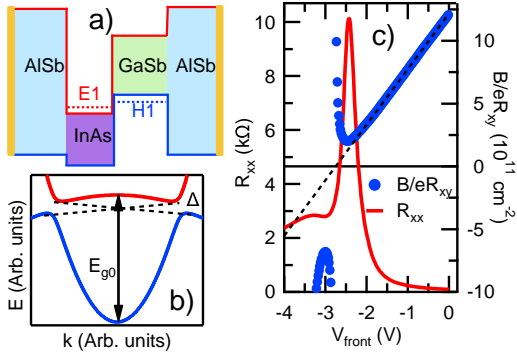}
\caption{\label{FIG. 1} {Panel a) shows energy spectrum of inverted CQW,
while band dispersion with linearly dispersing helical edges is shown
in panel b). Panel c) shows longitudinal resistance $R_{xx}$ (in
red) at $B=\unit[0]{T}$ and $B/eR_{xy}$ (in blue), taken at $B=\unit[1]{T}$,
vs. front gate bias $V_{front}$ for $\unit[50]{\mu m}\times\unit[100]{\mu m}$
device. As the $E_{F}$ is pushed into the hybridization gap $R_{xx}$
exhibits a strong peak, concomitantly $B/eR_{xy}$ becomes non-linear
signaling two-carrier transport and mini-gap entry. } }

\end{figure}

\begin{figure}
\includegraphics[scale=0.8]{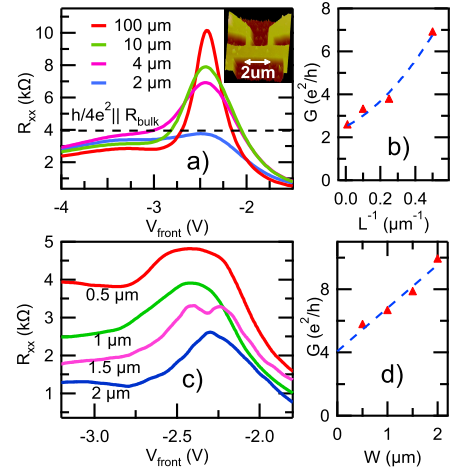}
\caption{\label{FIG. 2} {Panel a) shows $R_{xx}$ vs. $V_{front}$ for devices
with $\unit[L=100]{\mu m},$ $\unit[10]{\mu m},$ $\unit[4]{\mu m},$
and $\unit[2]{\mu m}$ (AFM image in inset) while $W$ is varied to
give constant geometric factor $\square=L/W=2$;$B=\unit[0]{T}$,
$T=\unit[300]{mK.}$ Resistance peaks decrease for shorter devices
and approach the limit $R_{bulk}||h/4e^{2}$ (dashed line) for $\unit[2]{\mu m}$
device. Panel b) shows gap conductance $G$ vs. $L^{-1}$ and is fitted
with Eq.\ref{eq:1} (dashed) giving coherence length $l_{\phi}=\unit[2.07\pm0.25]{\mu m}$.
Conductance difference between mesoscopic and macroscopic device is
$\sim4e^{2}/h$ suggestive of helical edge transport. Panel c) shows
$R_{xx}$ vs. $V_{front}$ for devices with $\unit[W=0.5]{\mu m},$
$\unit[1]{\mu m},$ $\unit[1.5]{\mu m},$ and $\unit[2]{\mu m}$;
$\unit[L=2]{\mu m}$. Resistance peaks decrease with increasing $W$.
Gap conductance $G$ vs. $W$ in panel d) shows linear relationship.
Intercept of the linear fit is $G_{edge}=\left(4.08\pm0.69\right)\frac{e^{2}}{h}$,
as expected for helical edge transport, while slope of the fit gives
bulk conductivity $g_{bulk}=\left(5.46\pm1.01\right)\frac{e^{2}}{h}$,
consistent with data in a). }}

\end{figure}

The experiments are performed on high quality 125Å InAs/50Å GaSb quantum
wells, in inverted regime. Sample fabrication and measurement details
are given in Ref. \citep{14,22}. Here the data were taken from eight
devices made from the same wafer. Fig. 1c shows longitudinal resistance
$R_{xx}$ (in red) vs. front gate bias $V_{front}$ of a Hall bar
with width $W=\unit[50]{\text{\textgreek{m}}m}$ and length $L=\unit[100]{\text{\textgreek{m}}m}$,
at $B=\unit[0]{T}$, $T=\unit[300]{mK.}$ As $V_{front}$ is swept
from $\unit[0]{V}$ to $\unit[-4]{V}$, $E_{F}$ is pushed from purely
electron to two-carrier hole dominated regime. When $n\sim p$, a
strong resistance peak of $R_{max}\sim\unit[10.2]{k\Omega}$ is observed,
which for this macroscopic sample reflects only the bulk transport,
with bulk gap conductivity of $g_{bulk}=\frac{\square}{Rmax}=5.05e^{2}/h$,
where $\square=L/W=2$. Entry into hybridization gap is also signaled
by non-linearity in $B/eR_{xy}$ (taken at $B=\unit[1]{T}$) shown
in Fig. 1c in blue. Negative values of $B/eR_{xy}$ indicate hole-dominated
regime although in two-carrier regime direct correspondence to carrier
density no longer exists. The size of the mini-gap can be determined
from the relative position in $V_{front}$ of the resistance dip,
which corresponds to the Van Hove singularity at the gap edge, and
the resistance peak which corresponds to the middle of the gap:\citep{11,14}
$\triangle=2\left(V_{peak}-V_{dip}\right)\frac{\Delta n}{\Delta V}\frac{1}{DOS},$
where $\frac{\Delta n}{\Delta V}$$=4.2\cdot10^{11}\mathrm{cm^{-2}/V}$
is the rate of carrier density change with $V_{front}$ and $DOS=(m_{e}+m_{h})/\pi\hbar^{2}$
is density of states, with carrier masses $m_{e}=0.03$ and $m_{h}=0.37$
(in units of free electron mass),\citep{11} giving $\triangle\sim\unit[4]{meV}$.
From the minimum in $B/eR_{xy}$ which corresponds to an anti-crossing
density of $n_{cross}\sim\unit[2\cdot10^{11}]{cm^{-2}}$, we can estimate
$E_{g0}=n_{cross}\frac{\pi\hbar^{2}}{m^{*}}\sim\unit[16]{meV}$, where
$m^{*}$ is the reduced mass. The expected bulk conductivity is then\citep{13}
$g_{bulk}\sim\frac{e^{2}}{h}\frac{E_{g0}}{\Delta}\sim\frac{4e^{2}}{h}$,
consistent with the observed value. 

\begin{figure}
\includegraphics[scale=0.9]{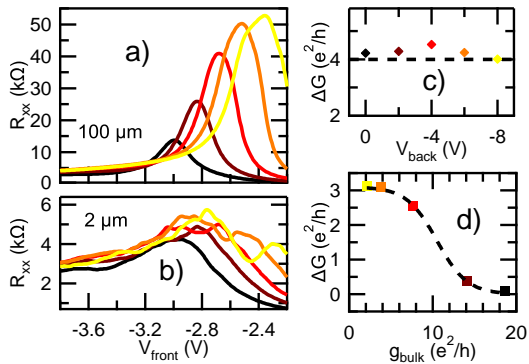}
\caption{\label{FIG. 3} {Panel a) shows $R_{xx}$ vs $V_{front}$ for devices
with $\unit[L=100]{\mu m},$ and in b) for $\unit[L=2]{\mu m}$ with
$V_{back}$ varied in $\unit[2]{V}$ steps from $\unit[0]{V}$ to
$\unit[-8]{V}$; $\square=2$, $B=\unit[0]{T}$, $T=\unit[20]{mK}$.
As $V_{back}$ is tuned to more negative values, the mini-gap moves
to smaller wave-vectors and the resistance peaks increase. The difference
in gap conductance between the $\unit[2]{\mu m}$ and $\unit[100]{\mu m}$
sample $\Delta G$ vs $V_{back}$ is shown in c), with $\Delta G\sim4e^{2}/h$
for all values of $V_{back}$. Note that $g_{bulk}\lesssim5e^{2}/h$.
Panel d) shows, $\Delta G$ vs $g_{bulk}$ for bias cooled sample
with larger bulk conduction. Edge conduction \textquotedblleft{}activates\textquotedblright{}
for $g_{bulk}\lesssim10e^{2}/h$. } }

\end{figure}

Fig. 2a shows resistance peaks for $L=\unit[100]{\mu m},$ $\unit[10]{\mu m},$
$\unit[4]{\mu m},$ and $\unit[2]{\mu m},$ with $\square=2$. The
resistance peak of the $L=\unit[100]{\text{\textgreek{m}}m}$ device
is used to estimate the bulk gap resistance $R_{bulk}\sim\unit[10.2]{k\Omega}$.
Surprisingly, a parallel combination of $R_{bulk}$ and the expected
edge resistance $h/4e^{2},$ gives a resistance value of $R_{bulk}||h/4e^{2}\sim\unit[3.95]{k\Omega}$
(dashed black line in Fig. 2a) which is just slightly above the measured
valued of $R_{max}\sim\unit[3.75]{k\Omega}$ for the $L=\unit[2]{\text{\textgreek{m}}m}$
device. A plot of the gap conductance $G$ vs. $1/L$ in Fig. 2b can
be fitted with Eq. \ref{eq:1}, obtaining $l_{\phi}=\unit[\left(2.07\pm0.25\right)]{\mu m}$
and giving further evidence for the existence of helical edge conduction
channels in mesoscopic samples. In fact, the difference in conductance
between the mesoscopic and the macroscopic samples is just slightly
above $4e^{2}/h$, as expected for helical edge modes.\citep{17} 

In width dependence experiment we fix $\unit[L=2]{\mu m}$, while
$W$ is varied from $\unit[W=0.5]{\mu m},$ $\unit[1]{\mu m},$ $\unit[1.5]{\mu m},$
to $\unit[2]{\mu m}$. While the resistance peaks shown in Fig. 2c
increase as $W$ is decreased, plot of $G$ vs. $W$ in Fig. 2d reveals
a reasonably linear relationship with an intercept of the linear fit
of $G_{edge}=\left(4.08\pm0.69\right)\frac{e^{2}}{h}$, in support
of helical edge transport. As an important check, the slope of the
same fit gives a bulk conductivity of $g_{bulk}=\left(5.46\pm1.01\right)\frac{e^{2}}{h}$
which is consistent with the value estimated earlier. Thus, both the
length and the width dependence of the gap conductance consistently
confirm the existence of helical edge channels in inverted InAs/GaSb
QWs.

\begin{figure}
\includegraphics[scale=0.85]{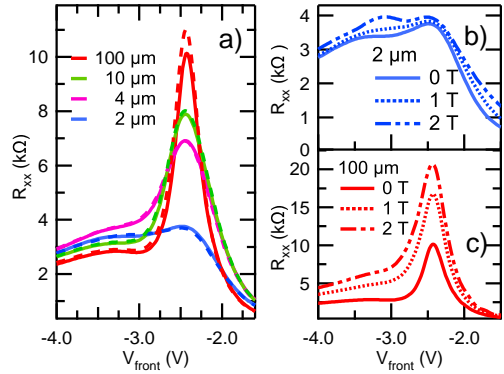}
\caption{\label{FIG. 4} {Panel a) shows $R_{xx}$ vs. $V_{front}$ at in-plane
field $B_{||}=\unit[0]{T}$ (full line) and $B_{||}=\unit[1]{T}$
(dashed line) for $\unit[L=100]{\mu m},$ $\unit[10]{\mu m},$ $\unit[4]{\mu m},$
and $\unit[2]{\mu m}$, indicating weak field dependence of gap resistance;
$T=\unit[300]{mK.}$Panel b) shows $R_{xx}$ vs. $V_{front}$ at perpendicular
fields of $B_{perpen.}=\unit[0]{T,}$ $\unit[1]{T,}$ and $\unit[2]{T}$
for $\unit[L=2]{\mu m},$ and in panel c) for $\unit[L=100]{\mu m}.$} }

\end{figure}

Using $V_{back}$ the anti-crossing point $k_{cross}$ can be tuned
to lower values, thereby supressing $g_{bulk}$. Fig. 3 shows $R_{xx}$
vs $V_{front}$ with $V_{back}$ varied in $\unit[2]{V}$ steps from
$\unit[0]{V}$ to $\unit[-8]{V}$ for devices of $L=\unit[100]{\text{\textgreek{m}}m}$
in a) and $L=\unit[2]{\text{\textgreek{m}}m}$ in b), $\square=2$
in both cases. As $V_{back}$ is tuned to more negative values, the
separation between the bands $E_{g0}$ is reduced, and the resistance
peaks of the $L=\unit[100]{\text{\textgreek{m}}m}$ sample increase
from $R_{max}\sim\unit[10]{k\Omega}$ at $V_{back}=\unit[0]{V}$,
to $R_{max}\sim\unit[50]{k\Omega}$ at $V_{back}=\unit[-8]{V}$. On
the other hand, the resistance peaks of the mesoscopic sample increase
only slightly, from $R_{max}\sim\unit[4]{k\Omega}$ at $V_{back}=\unit[0]{V}$,
to $R_{max}\sim\unit[6]{k\Omega}$ at $V_{back}=\unit[-8]{V}$. In
fact, the conductance difference between mesoscopic and macroscopic
samples, $\lyxmathsym{\textgreek{D}}G=G_{\unit[2]{\text{\textgreek{m}}m}}-G_{\unit[100]{\text{\textgreek{m}}m}}$
stays around $\sim4e^{2}/h$ for all values of $V_{back}$, as shown
in Fig. 3c, accounting for the helical edge transport. 

Data presented in Fig. 3a may suggest that edge conduction is completely
independent of gap bulk conductivity, $g_{bulk}$. However, this is
valid only in the regime of low $g_{bulk}$. Note that in Fig. 3a
$g_{bulk}\lesssim5e^{2}/h$. Using bias cooling technique,\citep{14}
the system can be pushed deeper into the inverted regime, i.e. a larger
$E_{g0}$ can be obtained, so that at $V_{back}=\unit[0]{V}$, $g_{bulk}\sim19e^{2}/h$,
while at $V_{back}=\unit[-8]{V}$, $g_{bulk}\sim e^{2}/h$. In this
case, the edge conductance, i.e. $\lyxmathsym{\textgreek{D}}G=G_{\unit[2]{\text{\textgreek{m}}m}}-G_{\unit[100]{\text{\textgreek{m}}m}}$,
goes from $\lyxmathsym{\textgreek{D}}G\sim0$ for large bulk conductivity
of $g_{bulk}\sim19e^{2}/h$ to about $\lyxmathsym{\textgreek{D}}G\sim3e^{2}/h$
as the bulk conductivity is reduced to $g_{bulk}\lesssim5e^{2}/h$,
as shown in Fig. 3d. The cut-off bulk conductivity at which edge conduction
\textquotedblleft{}activates\textquotedblright{} can be estimated
to $g_{bulk}\sim10e^{2}/h$. 

The apparent resilience of edge conduction to bulk transport is quite
unexpected, considering that a conductive bulk would allow edge electrons
to tunnel from one side to another, resulting in inter-edge scattering
and a reduced edge conductance.\citep{18,19} However, the inter-edge
tunneling probability may be significantly reduced by a large Fermi
wave-vector mismatch. The bulk gap states are inherited from the non-hybridized
band structure and have a Fermi wave-vector equal to $k_{cross}\gg0$
while edge modes, for $E_{F}$ situated in the middle of the gap,
have $k_{edge}\sim0$. Thus, due to $k_{edge}\ll k_{cross}$ edge
modes are totally reflected from bulk states. In fact, the tunneling
probability for the edge electrons will be proportional to the edge-bulk
transmission probability, which scales as $k_{edge}/k_{cross}$, as
well as the bulk transmission, which scales as bulk conductivity,
i.e. as $Ego\propto k_{cross}^{2}$. Hence, the overall inter-edge
tunneling probability will decrease as $k_{cross}$ is reduced, which
is in a qualitative agreement with data in Fig. 3d. Furthermore, due
to the low Fermi velocity of edge states $v=\frac{1}{\hbar}\frac{\partial E}{\partial k}\sim\frac{1}{\hbar}\cdot\frac{\Delta}{2k_{cross}}\sim\unit[3\cdot10^{4}]{m/s}$,
relativistic effects of Rashba spin-orbital interaction will be small,
and electron spins are expected to be aligned along the growth axis,
reducing inter-edge tunneling due to Pauli exclusion.\citep{20} 

The resistance peaks of mesoscopic samples show only a weak dependence
on in-plane and perpendicular magnetic fields, as shown in Fig. 4a
and 4b respectively, while macroscopic samples show a much stronger
dependence. At first glance, this appears to be in contrast to the
strong field dependence reported for HgTe/CdTe QWs\citep{7}. However,
even in HgTe a strong magnetic field dependence has never been observed
in the smallest micron size samples,\citep{20} but only in longer
$\left(\unit[20]{\text{\textgreek{m}}m}\right)$ samples.\citep{7}
In fact, it has been shown theoretically by Maciejko et al\citep{21}
that the magnetic field decay of edge modes depends sensitively on
disorder strength, with pronounced cusp-like features of magneto-conductance
occurring only when the disorder strength is larger than the gap.
The underlying physical picture is that by providing the states in
the bulk by large disorder, edge electrons can diffuse into the bulk
enclosing larger amounts of flux whose accumulation destroys destructive
interference of backscattering paths, resulting in a linear decay
of conductance with $B$. In the case of HgTe, large disorder was
provided by in-homogenous gating, which is more pronounced for longer
devices.\citep{7} 

In InAs/GaSb edge states are effectively decoupled from the bulk and
the above flux effect plays a lesser role, resulting in a weaker magnetic
field dependence of edge modes. However, the decay of bulk conductivity
itself with magnetic field may not necessarily be weak due to the
localization of non-hybridized carriers. This localization is more
pronounced for longer samples, which have stronger disorder. Thus,
longer samples are expected to show stronger magnetic field dependence,
as experimentally observed. We note here that localization at high
magnetic fields results in a dramatic re-entrant quantum Hall effect.\citep{22}
Such re-entrant behavior is a signature mark of topologically distinct
band structure,\citep{7} and its observation validates the topological
origin of helical edge modes at zero magnetic field. 

In conclusion, inverted InAs/GaSb CQWs in hybridization regime host
slowly-propagating helical edge modes which persist despite conductive
bulk and show only weak magnetic field dependence. This remarkable
property can be qualitatively explained by gap opening away from Brillouin
zone center, unlike in HgTe where the gap opens at $k\sim0$. Demonstrated
band structure tunability and good interface to superconductors make
this QSHI system a promising candidate in realization of exotic Majorana
modes.

The work at Rice was supported by Rice Faculty Initiative Fund, Hackerman
Advanced Research Program grant 003604-0062-2009, Welch Foundation
grant C-1682, and NSF grant DMR-0706634. I.K. acknowledges partial
support from M. W. Keck Scholar. We thank S.-C. Zhang for bringing
our attention to Ref. \citep{8}, S.-C. Zhang, X.-L. Qi, C. Liu, J.
Maciejko, and M. Konig for many helpful discussions.

\end{document}